\documentstyle[prb,aps,multicol]{revtex}
\input{epsf}
\begin{document}
\draft
\title{Localization in a rough billiard: A $\sigma$ model formulation}

\author{Klaus M. Frahm}

\address {Laboratoire de Physique Quantique, UMR 5626 du CNRS, 
Universit\'e Paul Sabatier, F-31062 Toulouse Cedex 4, France}

\date{18 December 1996, to appear in Phys. Rev. B, Rapid Communications}

\maketitle

\begin{abstract}
We consider the quantum dynamics of a particle in a weakly 
rough billiard. The Floquet operator for reflection at the 
boundary is obtained as a unitary band matrix. 
The resulting dynamics in angular momentum space can be treated in the 
framework of the one-dimensional supersymmetric 
nonlinear $\sigma$ model. We find analytically localization and 
the corresponding localization length $\xi=D_{cl}$ where $D_{cl}$ 
is the classical diffusion constant due to boundary scattering. 
\end{abstract}


\begin{multicols}{2}
\narrowtext

There is a close relation between random matrix theory \cite{mehta} and 
Efetov's supersymmetric nonlinear 
$\sigma$ model for disordered metals \cite{efetov}. Efetov calculated 
from the $\sigma$ model in the zero dimensional limit (energy scales 
smaller than the Thouless energy) the random matrix level statistics. 
Later Bohigas, Giannoni, and Schmit \cite{bohigas,leshouches} studied the 
level spacing statistics of chaotic billiards by numerical and semiclassical 
methods also confirming the random matrix behavior. 
Only quite recently, Muzykantskii et al. \cite{muzy} and 
Andreev et al. \cite{andreev} 
rederived the $\sigma$ model (containing the zero dimensional limit) 
for this type of systems. 

The chaotic billiards considered in Refs. [\ref{bohigas}-\ref{andreev}]
were characterized by one typical time scale $t_c$, the time between 
collisions with the boundary. There are also other type of chaotic 
systems in which a second time scale $t_D\gg t_c$ appears. $t_D$ is 
the time to cover the accessible phase space diffusively. A famous 
example is the kicked rotator \cite{rotator,shep} which exhibits diffusion 
in the momentum space (and localization in the quantum case). 
Aplying a new technic for averages over unitary matrices \cite{zirn_circ}, 
Altland and Zirnbauer were able to derive \cite{zirn_kick} 
the one-dimensional $\sigma$ model for this example. 
In this way, they obtained an analytical proof of localization and, 
for short time scales, also of diffusion in the kicked rotator model. 

Recently, the diffusive regime was also identified in two types of 
chaotic billiards: a nearly circular Bunimovich stadium billiard 
\cite{fausto} and slightly rough billiards with a general smooth boundary 
perturbation \cite{frahm_dima1}. In these examples, the classical 
diffusion constant in angular momentum space leads to an estimate of 
the localization length for the quantum case. If this length is below 
the classical angular momentum boarder at given energy, the classically 
chaotic billiard is no longer ergodic in the quantum case. Then the 
level spacing statistic changes from Wigner-Dyson to Poisson 
\cite{fausto,frahm_dima1}. 

In this paper, we consider as in [\ref{frahm_dima1}] a nearly circular 
billiard with a rough surface defined by the angle dependent radius 
$R(\theta)=R_0+\Delta R(\theta)$. 
The deformation is chosen as a function that depends only on a finite 
number of harmonics $\Delta R(\theta)/R_0=\mbox{Re}\ \sum_{m=2}^M \gamma_m\,
e^{i m \theta}$ where $\gamma_m$ are random complex coefficients. 
This type of billiard may have important physical applications in contrast 
to the stadium billiard which is of more theoretical interest. 
As examples, we can mention surface waves in water droplets which are 
practically static for light propagation \cite{droplet}, nonideal surfaces in 
microdisk lasers \cite{microdisk}, and capillary waves on a surface of 
small metallic clusters \cite{clusters}. 

The classical diffusive behavior due to boundary scattering is essentially 
determined by the roughness $\kappa(\theta)=(d R/d\theta)/R_0$ 
and the diffusion constant in angular momentum space is 
given by \cite{frahm_dima1} 
\begin{equation}
\label{eq1}
D_{cl}=4\,(l_{max}^2-l_i^2)\,\tilde \kappa^2\quad,
\end{equation}
with the angle average $\tilde\kappa^2=\langle \kappa^2(\theta)
\rangle_\theta$. $l_{max}=m R_0 v_F/\hbar$ is the classical angular momentum 
border at given velocity $v_F$ and $l_i$ is the initial angular momentum. 
The diffusion constant (\ref{eq1}) corresponds to an integer time 
variable $t$ measuring the number of collisions, a notion which we 
also adopt below. Using the analogy to the kicked rotator 
model \cite{shep} and numerical precise calculations of eigenstates, it was 
demonstrated \cite{frahm_dima1} that the localization length 
is given by $\xi=D_{cl}$. 

In this paper, we concentrate on an analytical approach for the quantum 
case which relies on certain simplifications but 
is otherwise not restricted in the parameter range as the direct numerical 
method. We will first derive an expression for the time evolution or Floquet 
operator corresponding to the reflection of the particle at the 
boundary. In the second part of the paper, we map the quantum 
dynamics of this operator onto the supersymmetric nonlinear 
$\sigma$ model. Here we will use a new technique which is quite different 
but presumably equivalent to Zirnbauer's method \cite{zirn_circ}. 

To investigate the quantum problem, we expand the wave function 
$\psi(r,\theta)$ with energy $E=\hbar^2 k^2/2m$ 
in terms of the Hankel functions which form a complete set:
\begin{equation}
\label{eq2}
\psi(r,\theta)=\sum_l \left(a_l\,H_{|l|}^{(+)}(kr)\,e^{il\theta} +
b_l\,H_{|l|}^{(-)}(kr)\,e^{il\theta} \right)\ .
\end{equation}
In principle, the regularity of $\psi$ at $r=0$ requires $a_l=b_l$ and the 
coefficients $a_l$ are therefore the amplitudes in angular momentum space. 
However, for conceptual reasons we disregard this condition and 
concentrate on the boundary condition $\psi[R(\theta),\theta]=0$. 
This condition results in a second equation $b_l=\sum_{l'} 
S_{ll'}(E)\,a_{l'}$ where $S_{ll'}(E)$ has the meaning of a scattering 
matrix for a wave packet reflected at the rough boundary. For 
time scales smaller than the collision time the particle does not 
see the origin $r=0$ and the dynamics are determined by an 
effective scattering problem with the ``Floquet operator'' $S(E)$. 
The energy eigenvalues 
of the full billiard are determined by the condition $\det[1-S(E)]=0$. 
To calculate $S(E)$, we multiply Eq. (\ref{eq2}) at 
$r=R(\theta)$ by $H^{(+)}_{|\tilde l|}[k R(\theta)]\,
e^{-i\tilde l\theta}$ and integrate over $\theta$. From this we obtain 
$S=-(A^{(-)})^{-1}\,A^{(+)}$ where $A^{(s)}$ ($s=\pm$) are matrices 
given by
\begin{equation}
\label{eq3}
A^{(s)}_{\tilde l,l}=\int_0^{2\pi} d\theta\ e^{i(l-\tilde l)\theta}
\,H^{(+)}_{|\tilde l|}[k R(\theta)]
\,H^{(s)}_{|l|}[k R(\theta)]\ .
\end{equation}
In the following, we will consider small values of the roughness, 
$\kappa\ll 1$, and therefore small boundary perturbations 
$\Delta R\ll R_0$. 
We also concentrate on the semiclassical limit $1\ll k\,\Delta R
\ll k\,R_0$, corresponding to a wavelength smaller than the geometrical 
length scales of the billiard. In this limit, there is an effective 
cutoff in the angular momentum space $|l|\le l_{max}\approx k R_0$. 
The modes $|l|>l_{max}$ correspond to evanescent modes which only give 
weak contributions. For the propagating modes $|l|<l_{max}$, we can use 
the quasiclassical approximation of the Hankel functions: 
\begin{equation}
\label{eq4}
H_l^{(\pm)}(kr) \approx 2\,[2\pi k_l(r) r]^{-1/2}\,\exp[\pm i 
(\mu_l(r)-\pi/4)]
\end{equation}
where $k_l(r)=k (1-r_l^2/r^2)^{1/2}$, $\mu_l(r)=\int_{r_l}^r\,d\tilde r\,
k_l(\tilde r)=k_l(r)\,r-|l|\,\arctan[k_l(r)\,r/|l|]$ and 
$r_l\approx |l|/k$ is the classical turning point. This approximation is 
excellent for all $r\gtrsim r_l$ in contrast to the standard asymptotical 
expression (which is only correct for $r\gg r_l$). The integral in 
Eq. (\ref{eq3}) can in principle be evaluated by a saddlepoint approximation. 
For the case $s=-$, we immediately obtain the diagonal approximation 
$A^{(-)}_{\tilde l,l}\approx \delta_{\tilde l,l}\,4/[k_l(R_0) R_0]$ whereas 
for $s=+$ there is for each value of $l-\tilde l$ a set of saddlepoints 
$\theta_s$ 
determined by $\tilde l-l=\{k_{l}[R(\theta_s)]+k_{\tilde l}[R(\theta_s)]\}
\,\Delta R'(\theta_s)\approx 2 \sqrt{l_{max}^2-l_i^2}\, \kappa(\theta_s)
= \sqrt{D_{cl}}\,(\kappa(\theta_s)/\tilde\kappa)$ 
if both $l$ and $\tilde l$ are close to some value $l_i$. This expression 
is just the classical map for one collision $l\to \tilde l$ given 
by the ideal reflection at the boundary at angle $\theta_s$ 
\cite{frahm_dima1}. For $|l-\tilde l|\gg \sqrt{D_{cl}}$ there are no 
saddlepoints $\theta_s$ and the matrix $A^{(+)}$ (and thus $S$) is an 
effective band matrix\cite{band_komp} of width $\sim\sqrt{D_{cl}}$. 
For the subsequent calculation, the semiclassical expression of $S$ in 
terms of the $\theta_s$ is not very convenient, and we keep the 
$\theta$ integral for $A^{(+)}$ instead. Expanding the phases $\mu_l$ for 
small $\Delta R$ and choosing values of $l$ and $\tilde l$ close to 
$l_i$, we finally obtain
\begin{equation}
\label{eq5}
S_{\tilde l,l}\approx e^{i\mu_{\tilde l}(R_0)+i\mu_l(R_0)+i\pi/2}\ 
<\tilde l| e^{i\,2\,k_{l_i}(R_0)\,\Delta R(\theta)}|l>.
\end{equation} 
This expression provides the first key result of this paper. Obviously, 
the matrix $S$ is unitary with a form similar to the Floquet 
operator of the kicked 
rotator if one replaces $\mu_l\to \tau\, l^2$ and $2\,k_{l_i}(R_0)\,
\Delta R(\theta)\to k_{\rm kr}\,\cos(\theta)$ where 
$\tau$ and $k_{\rm kr}$ are 
the standard kicked rotator parameters. Furthermore, we can consider 
the ``quantum'' diffusion constant defined by $D_q=\sum_{\tilde l}
(l-\tilde l)^2\,|S_{\tilde l,l}|^2$. In the limit $l_{max}\to\infty$, 
$D_q$ and $D_{cl}$ coincide due to the completeness of the angular 
momentum states $|l>$. 

The phases $\mu_l$ are rapidly varying with $l$ 
and in the regime of classical chaos they can be considered as quasi random 
\cite{phase_discuss}. We are thus led to the effective random matrix 
model $S_{\tilde l,l}=e^{i\mu_{\tilde l}+i\mu_l}\,U_0(l-\tilde l\,)$ 
with independent random phases \cite{phase_discuss2} $\mu_l$ 
and $U_0(l-\tilde l\,)=<\tilde l|\,
\exp[i\,f(\theta)]\,|l>$. The function $f(\theta)=\mbox{Re}\,\sum_{m=2}^M 
\alpha_m\,e^{im\theta}$ is a general kick potential with a finite 
number of harmonics [$\alpha_m=2\sqrt{l_{max}^2-l_i^2}\,\gamma_m$ for 
the rough billiard case (\ref{eq5})]. 
For later use, we introduce the notations $\bar f^2=\langle f(\theta)^2
\rangle_\theta=1/2\,\sum_m\,|\alpha_m^2|$ and $D=\langle f'(\theta)^2
\rangle_\theta=1/2\,\sum_m\,m^2\,|\alpha_m^2|$ for the typical 
value of $f$ and the ``diffusion'' constant $D$. In the following, we 
focus on the case $\bar f^2\gtrsim 1$ and 
consider the average probability $W(l,l_0;t)=\left\langle 
|<l|\, S^t\, |l_0>|^2\right\rangle_\mu$ for a transition $l_0\to l$ 
after a time $t$. 
The subscript $\mu$ denotes the average over the random phases. The 
Laplace transform of $W$ is given by
\begin{eqnarray}
\nonumber
&&\tilde W(l,l_0;\omega)  =  \sum_{t\ge 0} e^{i\omega t} W(l,l_0;t)\\
\nonumber
&&\quad = \frac{1}{2\pi}\int_{-\pi}^\pi dE
\Bigl\langle <l_0|(1+e^{-i(E-\omega/2-i0)}\, S^\dagger)^{-1}\,|l>\\
\label{eq6}
&&\qquad \times <l|(1+e^{i(E+\omega/2+i0)}\, S)^{-1}\,|l_0>\Bigr\rangle_\mu
\ .
\end{eqnarray}
The average over $E$ can be absorbed in the phase average and therefore 
we put $E=0$. 
In order to apply the supersymmetric technique, 
we write $\tilde W(l,l_0;\omega)=\partial_z
\partial_{z^*}\,F(z,z^*)\big|_{z=z^*=0}$ with the generating function:
\begin{equation}
\label{eq7}
F(z,z^*)=\left\langle \mbox{Sdet}^{-1}[1+e^{i(\omega/2+i0)}\, \hat U
+\hat J(z,z^*)]\right\rangle_\mu\ ,
\end{equation}
\begin{eqnarray}
\label{eq8}
\hat U &=&\left(\begin{array}{cc}
U &  \\
 & U^\dagger \\
\end{array}\right)\ ,\ 
\hat J(z,z^*)=\left(\begin{array}{cc}
J(z) &  \\
 & J(z)^\dagger \\
\end{array}\right)\ ,
\end{eqnarray}
$J(z)=z\,E_{l,l_0} \otimes P_{B},\ U=e^{2i\mu}\,U_0=e^{i\mu}\,S\,
e^{-i\mu}$. 
Here we have embedded the matrices in a super matrix space with 
two additional gradings, advanced-retarded grading (AR) and fermion-boson 
grading (BF). The blockstructure in (\ref{eq8}) corresponds to the 
AR-grading. $E_{l,l_0}$ is a matrix in $l$-space with only one 
nonvanishing entry $1$ at the position $(l,l_0)$. $P_{B}$ is the projector 
on the bosonic subspace and the notation for $U$, $U_0$, $S$, $e^{i\mu}$ 
refer to operators in the $l$ space. 
In contrast to conventional applications of the supersymmetry method, 
we are not dealing with Gaussian variables but with random phases instead. 
Recently, Zirnbauer introduced \cite{zirn_circ} for this case a new 
type of Hubbard-Stratonovich transformation 
which relies on interesting and also quite involved mathematical concepts. 

Here, we proceed differently by replacing the random phases in a suitable 
way with gaussian variables:
\begin{equation}
\label{eq9}
e^{2i(\mu_l+i0)}\ \to\ 1-2iA_l^\dagger\,(H_l+i0+iA_l\,A_l^\dagger)^{-1}
\,A_l
\end{equation}
where $H_l$ is an $N\times N$ gaussian random matrix (of the 
orthogonal ensemble in the limit $N\to\infty$) with variances 
$\langle H_{j,\tilde j}^2\rangle=
(1+\delta_{j,\tilde j})/N$. $A_l$ ($A_l^\dagger$) is an $N$-dimensional 
column (row) vector such that $A_l^\dagger\, A_l=1$ and $A_l\,A_l^\dagger$ 
is the projector on one particular state (e.g. $j=1$) in the Hilbertspace 
belonging to $H_l$. The expression on the r.h.s. of (\ref{eq9}) describes 
the $1\times 1$-scattering matrix of a chaotic 
cavity which is ideally coupled to one channel \cite{vwz}. It is well 
established \cite{lewenkopf,pietqdot} that the distribution of it 
(for the above given values of $A_l$ and $\langle H_{j,\tilde j}^2\rangle$) 
is given by the circurlar orthogonal ensemble which reduces to a 
uniformly distributed random phase in the $1\times 1$-case. 
The subscript $l$ denotes the fact, that the states $|j,l>$ belonging to each 
$H_l$ are by definition completely independent (orthogonal) for 
different values of $l$, i.e. the $H_l$ act on different $l$-sub blocks. 
In the following, we denote by $H$ the operator on the full set of states 
for all $l$ and $j$ which contains the $H_l$ in the diagonal $l$-blocks. 
In a similar way, we can arrange the operators $A_l$, ($A_l^\dagger$) 
to operators $A$ ($A^\dagger$) which perform transitions from the 
$U$-space to the $H$-space ($H$-space to $U$-space). 

The generating function can be expressed as a superdeterminant in the 
$H$-space:
\begin{eqnarray}
\nonumber
&& F(z,z^*)  =  \left\langle \mbox{Sdet}^{-1}_{(l,j)}[H+i0\Lambda + 
A \hat B(\omega;z,z^*) A^\dagger]\right\rangle_H\\
\label{eq10}
&&\qquad \qquad\quad\times \mbox{Sdet}^{-1}_{(l)}(1+\hat U_0+\hat J)\ ,\\
\label{eq11}
&& \hat B(\omega;z,z^*)  =  \left(\begin{array}{cc}
B(\omega,z) &  \\
 & B(-\omega,z)^\dagger \\
\end{array}\right)
\ ,\\
\nonumber
&& B(\omega,z) = i[1-e^{i\omega/2}\,U_0+J(z)]\,
[1+e^{i\omega/2}\,U_0+J(z)]^{-1}.
\end{eqnarray}
Here the subscripts $(l,j)$ or $(l)$ indicate the type of 
space over which the superdeterminant is taken. 
$\Lambda=\sigma_3^{(AR)}$ is the third Pauli matrix in the 
AR-grading. The ensemble average is now performed with respect to the 
$H_l$. In order to understand the physical meaning of this result, 
it is instructive to consider the matrix $\hat B$ at vanishing 
source term and frequency: $\hat B_0=\hat B(0;0,0)=i(1-U_0)/(1+U_0)
=\tan[f(\hat\theta)/2]$ which is hermitian and diagonal in the 
$\theta$-representation. The generating function (\ref{eq10}) corresponds 
to an $N$-orbital Hamiltonian with blockdiagonal diagonal GOE-entries 
and a coupling between the blocks given by $A\,\hat B_0\,A^\dagger$. 
For the moment, we consider the case of orthogonal symmetry 
$B_0^{\rm T}=B_0$ corresponding to $f(\theta)=f(-\theta)$. 
This case is actually not generic for general rough billiards. Below, 
we discuss the more generic unitary case. 

The technique to derive the supersymmetric nonlinear $\sigma$ model for 
this type of systems is well known \cite{iwz}. Performing the standard 
steps, we obtain:
\begin{eqnarray}
\label{eq12}
F(z,z^*) & = & \mbox{Sdet}^{-1}_{(l)}(1+\hat U_0+\hat J)\,
\int DQ\,e^{-{\cal L}[Q]}\ ,\\
\label{eq13}
{\cal L}[Q] & = & \frac{1}{2}\mbox{Str}_{(l)}\ln[\hat B(\omega;z,z^*)
+i\hat Q]\ .
\end{eqnarray}
Here $\hat Q$ is a block diagonal supermatrix (in $l$-space) 
with entries $Q(l)$ that 
are $8\times 8$-supermatrices with the nonlinear constraint $Q(l)^2=1$. 
The $Q(l)$ are taken from the space that corresponds to the 
$\sigma$ model with orthogonal symmetry \cite{efetov,vwz}. Omitting 
the details of the (standard) derivation of (\ref{eq12}), we mention however 
two points: First, the particular form of the operators 
$A$ and $A^\dagger$ allowed us to perform the trace over the $j$-quantum 
number on the level of the $\sigma$ model action. In this trace, 
for each $l$ only the particular state with $j=1$ contributes 
resulting in the action (\ref{eq13}). Second, the necessary saddlepoint 
approximation to arrive at the nonlinear constraint becomes indeed 
exact due to the limit $N\to\infty$. 

We are interested in the limit of long times ($t\gg 1$) or 
correspondingly of small frequencies ($\omega\ll 1$). Furthermore, 
the coupling term $\hat B_0$ favors slow fluctuations of the $Q(l)$ 
with $l$. We have therefore to apply a gradient expansion on the action 
with the first nonvanishing contribution:
\begin{displaymath}
\frac{1}{2}\mbox{Str}_{(l)}\ln(\hat B_0+i\hat Q)\simeq 
\frac{1}{8}\mbox{Str}_{(l)}\Bigl(\bigl((1+\hat B_0^2)^{-1}\,
[\hat B_0,\hat Q]\bigr)^2\Bigr)\ .
\end{displaymath}
We keep the linear order in $\omega$ and use $[g(\theta),Q(l)]\simeq 
i\,g'(\theta)\,\partial_l Q(l)$ for slowly fluctuating $Q(l)$ and some 
function $g(\theta)$. 
Going over to the continuous limit in $l$-space, we finally obtain the 
standard action for the one-dimensional $\sigma$ model
\begin{equation}
\label{eq15}
{\cal L}[Q]\simeq -\frac{1}{32}\int dl\Bigl(
D\,\mbox{Str}(\partial_l Q)^2+4i(\omega+i0)\,\mbox{Str}(Q\Lambda)
\Bigr)
\end{equation}
$+$ source term contributions where $D=\langle [f'(\theta)]^2\rangle_\theta(=
D_q=D_{cl})$ is the above introduced diffusion constant. 
Concerning the source term, we mention that the Sdet-prefactor in 
(\ref{eq12}) requires some care. A priori, it 
generates Greens functions of the type $(1+\hat U_0)^{-1}$ which 
correspond to the ``clean'' system without disorder and do not show any 
localization. However, a careful analysis shows that this 
contribution cancels with a corresponding term from the action. 

For the unitary case, we have to decompose $\hat B_0$ in even and odd 
contributions in $\theta$ resulting in $\hat B_0=\hat B_{0,\rm even}+
\tau_3 \hat B_{0,\rm odd}$ with $\tau_3=\sigma_3^{T}$ in the $T$-grading for 
the time reversel symmetry (this grading appears in addition to the $AR$- 
and $BF$-grading when deriving the $\sigma$-model). Assuming that 
the odd term is comparable to 
the even term, we obtain in the action the extra term 
$\sim \int dl\,\mbox{Str}([Q,\tau_3]^2)$. For slowly fluctuating 
$Q(l)$, this term produces massive modes in the offdiagonal $T$-blocks of 
$Q$. These ``cooperon'' modes are therefore suppressed and $Q$ becomes 
$T$-block diagonal \cite{efetov}. As a result, we obtain the 
unitary $\sigma$ model 
\cite{unit_discuss} 
with the same action (\ref{eq15}).

It is well known that the $\sigma$ model (\ref{eq15}), 
gives rise to diffusive dynamics with the diffusion constant $D$ for 
sufficiently short length scales. For long length scales it 
provides as in quasi 1d disordered wires \cite{larkin,efetov} 
localization with the localization length $\xi=\beta D/2$ where 
$\beta=1$ ($2$) is the symmetry index for orthogonal (unitary) symmetry. 
The localization length for $\beta=2$ 
agrees with the numerical findings of Ref. [\ref{frahm_dima1}]. 

The above used gradient expansion is well justified for the limit 
$\bar f^2 \gg 1$. It is an interesting question in how far the action 
(\ref{eq15}) remains correct for $\bar f^2 <1$. The answer is indeed 
not obvious, since we may still have $D\gg 1$ for $M\gg 1$. 
Now we can apply on (\ref{eq13}) an expansion for small $B_0\simeq 
(1/2)\, f(\theta)$. The ``kinetic'' part of the action is then given by 
a term $\sim\sum_{l,\tilde l} |\alpha_{|l-\tilde l|}|^2\,
\mbox{Str}[Q(l)\,Q(\tilde l)]$. This contribution couples the $Q$-fields 
on a strip of width $M$. The excitation ``energy'' due to 
$Q(l)$ fluctuations with a wavelength $M$ (the ``smallest'' nonzero mode)
is estimated as $\sim M (D/M^2) \sim D/M$. The gradient expansion is 
justified if this mode gives only a small contribution, i.e. if it 
becomes massive for $D\gg M$. For $D\ll M$, the $Q(l)$ fields may fluctuate 
nearly independently and we enter a regime in which the eigenphases and 
eigenfunctions of the matrix $S$ can be calcuated by simple perturbation 
theory due to weak coupling. 

In [\ref{frahm_dima1}], a model with harmonics of different weights 
$\alpha_m\sim 1/m$ was considered such that $D\sim M \bar f^2$. 
Here, the two conditions $\bar f^2> 1$ and $D > M$ coincide and 
we have only two regimes of weak and strong coupling between different 
$l$-states. However, we may also study \cite{frahm_dima1} 
a model with harmonics of equal 
weight, $\alpha_m\sim const.\,$. 
Then we have $D\sim M^2 \bar f^2$ and 
there is a third nontrivial regime charaterized by $M< D < M^2$ where the 
$\sigma$ model action (\ref{eq15}) is valid and the typical coupling 
is still small. This regime correponds for $S$ to a band matrix with a 
strong diagonal in the so-called Breit-Wigner regime \cite{frahm_dima1,shep2}. 

In summary, we have derived the Floquet operator in angular 
momentum representation for a weakly rough billiard. This operator 
describes the reflection of a quantum particle at the boundary. 
In the chaotic regime, the dynamics can be mapped on a 
supersymmetric nonlinear $\sigma$ model in the one dimensional 
angular momentum space. As a result, 
we obtain analytically diffusion with the diffusion constant 
$D_{cl}$ [Eq. (\ref{eq1})] for short time scales and localization 
with the localization length $\xi=D_{cl}$. 
A generalization to three dimensional billiards is certainly 
an important and interesting topic of future research. 

We have also introduced an alternative way to Zirnbauer to deal with 
the random phases. This technique relies on well known relations 
between the circular and gaussian random matrix ensembles and can 
also be applied to random unitary matrices of arbitrary dimension. 
In analogy to Altland and Zirnbauer, it is also possible only to average 
over the quasi energy as indicated in (\ref{eq6}). 
In this case, the matrix 
$\hat Q$ in (\ref{eq12}) will be nonlocal in $l$ and the subsequent 
expansion of the action will be more involved \cite{zirn_kick}. However, 
we do not expect any serious modifications from this. 
Especially, we recover for $f(\theta)=k_{\rm kr}\cos(\theta)$ the asymptotic 
expression $\xi=k_{\rm kr}^2/4$ for the localization length of the quantum 
kicked rotator \cite{rotator,shep,zirn_kick}. 

The author acknowlegdes D. L. Shepelyansky for fruitful and 
inspiring discussions.

\end{multicols}


\begin{thebibliography}{99}
\bibitem{band_komp} Strictly speaking this is only valid in the limit 
	$D\gtrsim M^2$. Otherwise the effective bandwidth is $\sim M$ due 
	to small but direct transitions.
\bibitem{phase_discuss} For a finite number $M$ of harmonics $\gamma_m$ 
	there is a chaos border \cite{frahm_dima1} 
	 $\kappa_c$ for the roughness which vanishes 
	very rapidly with increasing $M$ ($\kappa_c\sim 4M^{-5/2}$ for a 
	model with $\gamma_m\sim 1/m$). 
	The random phase approximation only applies for $\kappa>\kappa_c$.
\bibitem{phase_discuss2} For simplicity, we disregard here the 
	symmetry $\mu_l=\mu_{-l}$ leading to the Shnirelman peak 
	at $s=0$ in the level spacing distribution \cite{frahm_dima1}.
\bibitem{unit_discuss} It is also possible to choose in 
	(\ref{eq9}) a gaussian matrix $H_l$ of the unitary ensemble. 
	This gives directly the $\sigma$ model for the unitary case.
\end{thebibliography}
\end{document}